# Redactable Blockchain Solutions for IoT: A Review of Mechanisms and Applications

Arpish R. Solanki

## Abstract

The integration of blockchain technology with the Internet of Things (IoT) presents a promising solution to enhance data security, integrity, and trust within IoT ecosystems. However, the immutable nature of blockchain technology conflicts with data redaction requirements mandated by data protection laws. This paper provides a comprehensive review of the current state of redactable blockchains and redaction mechanisms, particularly focusing on their application within IoT contexts. Through an extensive review of existing literature, this paper identifies key challenges and opportunities in implementing redactable blockchains for IoT data management. Various redaction mechanisms are explored, and the paper examines IoT implementations and use cases where redactable blockchains are employed to address data protection concerns.

## 1. Introduction

The global Internet of Things (IoT) market is expected to grow to $4,062.34 billion by 2032 (Fortune Business Insights, 2024). The IoT has experienced significant growth with billions of connected devices worldwide (Fuller et al., 2020). Statistics show that there were 21.7 billion active connected devices worldwide, including 11.7 billion IoT connections (Velayudhan et al., 2022). By 2025, over 40 billion IoT devices are expected to be in use and generate huge amounts of data annually (Gupta and Ghanavati, 2022). This rapid expansion required the introduction of strict data protection regulations aimed at protecting individual privacy and ensuring secure handling of data.

IoT connects billions of smart devices, enabling data collection and intelligent decision-making in diverse applications, from smart homes to industrial automation. At the same time, blockchain technology, with its secure, immodifye and decentralized ledger system, has emerged as a revolutionary force capable of transforming industries through greater security, transparency and efficiency (Islam et al., 2023). These properties open a new domain for IoT applications (Mahlous and Ara, 2020). The fusion of blockchain technology with IoT networks has become a focus of contemporary research and offers solutions to pressing security concerns (Pathak, Al-Anbagi, and Hamilton, 2023).

However, the immutability of blockchain technology presents a significant challenge when it comes to balancing data flexibility with legal requirements. Although immutability guarantees the security and integrity of data stored on the blockchain, it can create obstacles when data redaction is required for regulatory compliance, especially in cases involving personal data (Politou et al., 2021). This conflict may hinder the widespread adoption of blockchain technology. Data redaction on the IoT blockchain is essential to comply with relevant laws and regulations that require correction and deletion of data subjects' information. Recent efforts have been made to address the tension between the immutability of blockchain and the need for data redaction.

This tension between the need for data flexibility and the fundamental principles of blockchain forms the core focus of this review. Redaction mechanisms, particularly in IoT contexts, offer a promising avenue to meet both technological and legal requirements. This paper explores the integration of IoT and blockchain technologies, specifically addressing the redaction requirements necessitated by data protection laws.

The main objective of this review is to analyse the existing literature on IoT and blockchain integration, with a particular focus on redactable blockchains and redaction mechanisms. Our goal is to identify and evaluate these mechanisms, assess their applicability to IoT-specific use cases, and highlight current challenges and possible future research directions. Given the early stages of the field, our review includes a wide range of scholarly articles and case studies published in the last decade.

The paper is organized as follows: We begin with an examination of data protection laws across various jurisdictions, focusing on requirements for data deletion, consent, rectification, and objection. Following this, we discuss the integration of IoT and blockchain technologies, highlighting existing solutions and categorizing them based on previous research. After detailing their integration, we explain how the immutability of blockchain can create challenges. We then examine redactable blockchains, explaining the overall concept and identifying specific redaction mechanisms. Note that in this paper, the terms "mutable blockchain" and "redactable blockchain" will be used interchangeably. We analyse IoT-specific implementations and use cases of redactable blockchains. Finally, we identify the challenges faced by these methods and implementations and suggest possible future research directions before concluding the paper.

## 2. Data Protection Laws and Regulatory Requirements

The increasing volume of data, coupled with the advent of Big Data, has introduced significant challenges in ensuring privacy protection and regulatory compliance. The enforcement of data protection regulations has become more prominent, granting users unprecedented control over their personal data and requiring businesses to adhere to strict privacy and security standards.

Multiple jurisdictions have established data protection and privacy laws that can be categorized based on their similarities and relevance. These laws include rights such as the right to rectification, the right to erasure, the right to restrict processing, and the right to object. The common thread among these regulations is the emphasis on protecting data privacy and respecting the rights of individuals in relation to their personal data.

## 2.1. Right to Rectification

Article 8 of the Charter of Fundamental Rights states that every individual has the right to protect their personal data. It also grants individuals the right to rectify inaccurate personal data or complete missing information. This principle is supported by various regulations, including:

- CCPA 1798.106
- GDPR Article 16
- FADP Article 32
- PIPL Article 46
- HIPAA 45 CFR 164.526
- DPDP Section 12
- UK GDPR Article 16 and Article 5(1)(d)
- DPA 2018 Section 46

These laws and regulations emphasize the importance of data accuracy, transparency and security. They give individuals the right to request corrections or rectifications to ensure data accuracy and control over their information. Individuals can submit rectification requests either in writing or verbally, and organizations must respond formally within a defined timeframe. However, organizations can refuse a rectification request under specific circumstances.

## 2.2. Right to Restrict Processing

Closely linked to the right to rectification and the right to object, individuals have the right to request the restriction or suppression of their personal data. This right is not absolute and applies only under specific conditions. It is supported by:

- GDPR and UK GDPR Article 18
- PIPL Article 44

- DPDP Section 6(4)
- FADP Article 32

These regulations highlight the importance of individual control over personal data. Although the details may vary, the basic principle is consistent: individuals can limit the processing of their data under certain conditions, and organizations must comply with such requests.

Individuals also have the right to object to the processing of their personal data for specific purposes, such as direct marketing or profiling activities. This right ensures that individuals can control how their personal data is used, particularly for activities they may find intrusive or objectionable.

### 2.3. The Right to Be Forgotten Across Various Jurisdictions

Individuals have the right to request the erasure of their data, supported by various regulations worldwide. This right allows individuals to request the deletion of their personal data from data controllers, particularly when the data is no longer needed for its original purpose or when consent for its processing is withdrawn. Requests can be made verbally or in writing, and organization are typically required to respond within defined timeframe. However, this right is not absolute and applies only under specific conditions. This principle is supported by various regulations, including:

- GDPR Article 17 and Article 5(1)(d) - European Union
- CCPA Section 1798.105 - California
- UK GDPR Article 17 and Data Protection Act 2018, Section 47
- PIPL Article 47 - China
- DPDP Section 12(3) - India
- FADP Articles 32 - Switzerland

In California, under the CCPA, residents can request the deletion of their personal information held by businesses. Businesses must comply within 45 days unless certain exceptions apply, such as the necessity to complete a transaction, detect security incidents, or comply with legal obligations.

The UK GDPR, supplemented by the Data Protection Act 2018, provides a similar right to erasure as the EU GDPR. Individuals can request the deletion of their personal data if it is no longer necessary, if consent has been withdrawn, or if the data was processed unlawfully. Organizations must respond within one month.

Under China's PIPL, individuals can request the deletion of their personal data when the purpose of data processing has been achieved, consent is withdrawn, or the processing violates laws or agreements. Data controllers must delete the data without undue delay upon request.

The DPDP act in India grants individuals the right to request the deletion of personal data if it is no longer necessary, if consent has been withdrawn, or if the data was processed unlawfully. Organizations are required to comply with these requests promptly.

The FADP in Switzerland allows individuals to request the deletion of their personal data if it is no longer necessary for the original purposes, if consent has been withdrawn, or if the data was processed unlawfully. Data controllers must act on such requests without undue delay.

### 2.3.1. Case Studies: Google Spain and Subhranshu Rout:

In the landmark Google Spain case (Judgment of the CJEU in Case C-131/12), Europe's highest court ruled that Google must remove certain search results relating to a Spanish national, highlighting that search engines are data controllers. This case underscored that while search engine results may be removed, the underlying content on third-party websites may remain if those sites have legitimate grounds for processing the information.

Similarly, in India, the Orissa High Court in the Subhranshu Rout case examined the right to be forgotten, emphasizing the victim's privacy rights and referencing international standards like the GDPR. The court noted the impracticality of victims approaching the court each time they needed data erased from social media platforms, pointing to the need for data controllers to act responsibly and promptly.

These deletion rights across various jurisdictions reflect a global recognition of the importance of individual control over personal data. While the specifics can vary, the underlying principle remains the same: individuals have the right to request the deletion of their personal data under certain conditions, and organizations are obligated to comply within stipulated timeframes. As the adoption of blockchain technology and IoT systems continues to grow, the need for effective data redaction mechanisms to ensure data privacy and compliance with evolving regulations becomes more apparent.

## 3. Integration of IoT and Blockchain Technologies

The Internet of Things (IoT) is a critical and progressive technology characterized by recent remarkable advances. It extends across different areas and underlines its diverse relevance. However, despite its growing potential, the IoT landscape is fraught with numerous challenges, particularly when it comes to the security and integrity of IoT systems. At the same time, efforts continue to be made to mitigate these challenges and strengthen the effectiveness of IoT infrastructures. Blockchain technology is proving to be a standout solution known for its ability to maintain security and integrity autonomously and without central oversight. As a viable solution to certain problems in the IoT space, blockchain provides mechanisms to overcome security and privacy deficiencies. Its decentralized architecture facilitates data integrity and transaction transparency, thereby avoiding the need for centralized authorities. By eliminating the vulnerabilities of centralized IoT frameworks, blockchain integration improves the overall security posture of IoT networks. The immutable nature of the blockchain ledger ensures a tamper-proof and verifiable transaction history and underpins the credibility of IoT applications. Additionally, the blockchain's distributed storage paradigm ensures network integrity even in the event of a node compromise due to redundant data copies. Consequently, this resilience leads to greater operational efficiency compared to traditional methods. Despite the mutual benefits that arise from their integration, it is imperative to recognize the inherent differences between blockchain and IoT technologies. While blockchain requires significant computing resources, the Internet of Things is based on ubiquitous objects with limited computing capabilities. Additionally, IoT requires low latency, an aspect where blockchain has latency in transaction throughput in comparison.

Existing research aims to address and improve IoT challenges. Most blockchain IoT integration systems attempt to replace existing components or mechanisms of traditional IoT systems to achieve improvements. In recent years, the focus has been on improving security and integrity (Tran, Babar, and Boan, 2021).

Various solutions have been proposed, each with different architectures and methods for integrating blockchain into existing IoT systems. To understand where blockchain is physically implemented in these solutions, we will use the edge-fog-cloud model to distinguish blockchain positions (Tran, Babar, and Boan, 2021).

- *Cloud layer:* This layer is globally accessible and includes computing resources located outside of IoT devices and edge devices. The most common deployment location for blockchain is in the cloud, typically using public blockchains or blockchain-as-a-service components. Due to the availability of sufficient computing resources, this layer generally includes full nodes.
- *Fog layer:* The fog layer acts as an intermediate bridge between the cloud and the edge devices, extends cloud services to the edge and facilitates intermediate processing with its computing capabilities. Another common deployment location for blockchain is the fog layer.
- *Edge layer:* This layer consists of end devices used for low latency processing close to the data source. Deploying blockchain at the edge layer is less common and difficult due to the limited computing resources and storage capacity of edge devices.

Solutions often use both full nodes and lightweight nodes in blockchain-integrated IoT systems. The typical approach is to place full nodes in the cloud and lightweight nodes in deeper layers such as the fog or even at the edge. Some solutions also use multiple ledgers, with different sets of nodes managing these multiple ledgers.

## 3.1. Common Models and Approaches

Many solutions have been proposed for IoT challenges that share similar patterns and basic design principles. Based on existing surveys, the most common model types are the following:

*M2M Autonomous Trading:* Machine-to-Machine (M2M) trading in the IoT context, enhanced by blockchain technology, enables automated information and transaction exchanges among devices without human intervention (Gong, Liu, and Wang, 2020). Devices autonomously interact, share data, and execute predefined logic, while blockchain provides a secure, decentralized ledger for recording all transactions. Smart contracts enforce terms autonomously (Zhang and Wen, 2016), making this model suitable for applications like supply chain management and energy trading, where secure, autonomous device transactions are essential.

*Authentication Management:* Numerous studies have focused on blockchain-based authentication systems for IoT. Unlike traditional authentication systems that rely on a centralized entity (Shen et al., 2020), which can result in high key management overhead or reliance on a trusted third party, this approach leverages peer-to-peer authentication by storing device descriptions and identities in the blockchain (Zhang et al., 2020). The on-chain data manages access control policies (Putra et al., 2021; Ding et al., 2019; Kumar et al., 2023) and verifies identity claims without relying on a central authority.

*Blockchain-based access control in IoT:* Access control is critical in IoT systems, and traditional systems often rely on centralized, trusted third parties. Blockchain is used in access control mechanisms for IoT systems by encoding access rules and conditions into smart contracts. When an IoT device requests access, the smart contract validates the request based on predefined rules (Namane and Dhaou, 2022). On-chain data includes policies (Dukkipati, Zhang, and Cheng, 2018), and consensus ensures agreement on access control decisions.

*Firmware update distribution:* A common use case is the secure and distributed firmware update mechanism for IoT using blockchain. This approach preserves the characteristics of the CIA triad (confidentiality, integrity, availability) (Arbabi and Shajari, 2019). Manufacturers or providers add metadata, hashes or update instructions to the blockchain. Nodes usually participate voluntarily and the immutable nature of the blockchain prevents tampering with updates. In many cases, distribution nodes are incentivized by manufacturers or providers (Leiba et al., 2018). Blockchain serves as a delivery infrastructure or process coordinator, enforcing agreements between update producers and consumers. Some solutions also use third-party distributed databases and store only hash values of update image blocks (Choi and Lee, 2020) on-chain to verify integrity and manage availability.

*Blockchain as a secure computing environment:* Another approach uses blockchain as a secure computing environment (Park and Kim, 2017) on untrusted computing nodes in IoT networks. This leverages the programming capabilities of smart contracts on blockchains. Blockchain prevents unintended operations in the infrastructure, improves security and integrity, and protects the network.

*Blockchain for monitoring and managing IoT devices:* This approach stores configuration changes on the blockchain (Košťál et al., 2019), ensuring the propagation of configuration changes to IoT devices. It leverages the immutability of blockchain to prevent unauthorized changes to IoT device configurations (Huh, Cho and Kim, 2017), storing change histories (Ramane et al., 2021) for easy incident recovery. Chain Code logic includes CRUD (create, read, update, delete) operations on on-chain data, which in this case relates to IoT device configurations.

*Blockchain as a Trust Rating Repository for IoT Networks:* IoT networks are vulnerable to network attacks by malicious users. This method dynamically evaluates the behavior and reputation of participants in IoT systems. It stores the reputation or trust ratings of IoT participants on the blockchain. Some solutions use on-chain logic to calculate ratings, but most prefer off-chain calculations. Many notable edge blockchain integration models leverage this approach by storing vehicle reputation (Kang et al., 2019; Yang et al., 2019; Singh and Kim, 2018) at the edge level or on vehicle computers (Yang et al., 2017).

### 3.1.1. Blockchain for Storing Sensor Data:

Blockchain is inherently immutable and transparent, making it ideal for ensuring accurate recording and tracking of data generated by IoT devices (Baracaldo et al., 2017). Integrating blockchain into IoT systems can ensure traceability from source to destination. This solution is beneficial not only for the data generated, but also for entities, their identities and policies. It creates a cross-organizational provenance chain to verify accuracy, accountability and non-repudiation (Liang et al., 2017). Common approaches include using blockchain as a communication channel or as secure data storage.

*Blockchain as a communication channel or intermediary:* Blockchain serves as a distributed, immutable record of transactions in a secure and tamper-proof manner. In this approach, the blockchain is used to store the data collected by IoT devices, and all legitimate participants in the network access it as a common communication channel or as a verifying intermediary for communication. The decentralized and tamper-proof ledger increases the security and trust of data exchange within and between IoT networks. Here, the blockchain acts as an immutable record keeper and workflow orchestrator, working without intermediaries (Albulayhi and Alsukayti, 2023). Smart contract logic includes transaction rules (Kerrison, Jusak and Huang, 2023), and on-chain data can include instructions for devices, transaction records, IoT-generated data, or hashes of these data.

*Blockchain as secure data storage:* In this solution, the blockchain acts as a tamper-proof record of data entries or data hashes. Hashes serve as pointers or indexes for off-chain data storage. Blockchain maintains the integrity and security of data generated by IoT devices. Due to its immutability, data once stored on the blockchain is difficult to alter and provides an auditable view to trace its origin. Blockchain protects data at rest or hashes/indexes of that data, as well as event records from IoT systems.

However, this approach faces challenges. As the number of IoT nodes increases, blockchain performance and storage capacity are strained. Blockchain requires every full node in the network to store a copy of the ledger, and it struggles to process and validate the high number of transactions required, potentially affecting the responsiveness of the IoT. Additionally, the use of incentive networks such as Ethereum, which is commonly employed, brings challenges such as transaction fees or costs.

Scalability solutions include dividing the blockchain into subsections, using multiple chains (Maftei et al., 2023), or a hierarchical structure of multiple chains that allows transactions to be processed in parallel (Hafid, Hafid and Samih, 2020). Another common approach is to use hashes or data indexes. Hash-based methods generate unique identifiers for each data packet (Tekchandani et al., 2023b;Zyskind, Nathan and Pentland, 2015; Ali, 2017), with hash values serving as pointers to actual data stored off-chain (Ayoade et al., 2018). This method overcomes capacity issues but has some security drawbacks. Some solutions combine on-chain and off-chain storage by using Distributed Hash Tables (Maymounkov and Mazières, 2002; Zyskind, Nathan and Pentland, 2015), InterPlanetary File System (IPFS), or cloud storage (Liang, Zhao, et al., 2017).

Encrypting data and applying access control are also common approaches to solving security problems. Despite the challenges, many solutions propose storing data directly on the blockchain. Although storage solutions for large amounts of data are less common, they remain a viable option for IoT systems.

## 4. Challenges of Immutability in IoT and Blockchain Integration

Recording IoT data on the blockchain is done through a distributed ledger system, where data is stored in a series of blocks, each containing multiple transactions. These blocks are linked together using cryptographic hashes, forming an immutable ledger. A new block is added to the ledger through a consensus between nodes. This ledger is replicated and shared among multiple nodes within the network.

Traditional blockchains, first introduced by Bitcoin and later Ethereum, are based on the principles of immutability and decentralization (Attaran and Gunasekaran, 2019). These blockchains operate without a central authority and distribute control across multiple nodes in the network. Once data is recorded on the blockchain, it cannot be changed. Changing past transactions would require changing all subsequent blocks.

However, these fundamental features of blockchain can present significant challenges. The inherent immutability of blockchains means that once data is recorded, it cannot be changed. This poses a problem in IoT systems that generate large amounts of data. Individual privacy becomes an important issue because immutability can prevent data from being deleted or modified, which conflicts with data protection regulations that give individuals the right to delete their data or set retention periods for stored data. In addition, individuals can revoke their consent to data collection and processing and request the deletion of previously collected data. Traditional blockchains struggle in such scenarios because modifying recorded data requires extensive computing resources, which contrasts with the limited computing capacity of IoT devices.

The Distributed Hash Table (DHT) used in the InterPlanetary File System (IPFS) lacks inherent mechanisms to remove data once it is added to the network. Data persists on the IPFS network as long as at least one node continues to share and pin it (Politou et al., 2020), even if it is deleted by a single node. Cached files may remain accessible on other nodes, especially if they are pinned, maintaining their availability. While IPFS does not provide efficient methods for deleting illegal or copyrighted content, it does allow private networks to be built via IPFS clusters. In such networks, one entity controls all peers, enabling shared resolution and deletion of files. However, the complete removal of content from the public IPFS network is uncertain.

## 5. Redactable Blockchains and Redaction Mechanisms

Redactable blockchains offer a promising solution to the challenges of data modification and deletion in blockchain-based IoT systems. These blockchains provide controlled flexibility and allow data entries to be changed or removed without affecting the sequential integrity of the chain. This feature enables compliance with privacy regulations while maintaining the effectiveness of the blockchain. Integrating redactable blockchains into IoT solutions can improve data management by balancing integrity, security and legal compliance.

A notable historical example of blockchain redaction is the DAO hack in 2016. The DAO was a smart contract on the Ethereum blockchain that raised over $150 million worth of Ether. Due to a vulnerability in the DAO's code, attackers managed to steal a third of the funds. To mitigate this, the Ethereum community performed a hard fork that rolled back transactions and led to the creation of two separate blockchains: Ethereum and Ethereum

Classic. Another early effort to modify blockchain transactions was Reversecoin (Challa, 2020), which introduced "Reversible Transactions." This feature allowed users to reverse transactions within a specific timeout period. Reversecoin had two types of accounts: Standard Accounts, which functioned like Bitcoin, and Vault Accounts, which were backed by both online and offline keys. If an unauthorized transaction has occurred, users can reverse it using the offline key during the timeout period. Despite its innovative approach, Reversecoin did not gain widespread adoption.

## 5.1. Redaction Mechanisms

Past research has explored various mechanisms to enable data modification or deletion on blockchain systems. In this section, we will examine these approaches in detail.

### 5.1.1. Chameleon Hash

The predominant approach to achieve redaction in blockchain technology relies heavily on the use of the Chameleon hash method (Ali, Yusoff and Hasan, 2023). Unlike traditional hash functions, where every change in input always results in a unique hash value, Chameleon Hash represents a cryptographic hash function that can generate a trapdoor key. This key can be used along with the hash value to derive a new input that yields the same hash (Krawczyk and Rabin, 1998). The trapdoor key can be used to find collisions and enable data changes. Conversely, without the trapdoor key, the chameleon hash works like any typical collision-resistant hash function, so it is not computationally possible to generate the same hash for different inputs. Consequently, it becomes virtually impossible to change input data without affecting the hash.

In blockchain systems, these chameleon hash functions replace traditional hash functions such as SHA256 for linking blocks. This modification is intended to protect the integrity of the blockchain, as only those who have access to the Trapdoor key can manipulate the data. Additionally, many solutions use "scars" as indicators to show that data has been changed (Politou et al., 2021).

Ateniese et al. (2017) proposed the first solution based on standard chameleon hash, integrating chameleon hash functions, public key cryptography and non-iterative zero-knowledge proofs. This approach modifies the block structure to include a field for recording the randomness "r", which only the holder of the trapdoor key can compute. The solution grants power to a central authority or shares the trapdoor key with a predefined group of participants who can derive the full key through multiparty computation. Data deletion occurs at the block level.

Derler, Samelin, et al. (2019) introduced fine-grained data reduction control through a policy-based chameleon hash function combined with ciphertext policy attribute-based encryption (CP-ABE). This method requires two trapdoors for collision: a standard trapdoor and a ephemeral trapdoor specified during hashing. To compute the hash according to the access policy, a user generates a chameleon hash using the ephemeral trapdoor and encrypts it with CP-ABE. To modify the transaction, a user with access reconstructs the ephemeral trapdoor to compute the collision for the transaction hash. Redaction occurs at the transaction level, with transactions categorized as modifiable or non-modifiable.

Tian et al. (2020) proposed a policy-based chameleon hash with black box accountability (PCHBA) that enables transaction creation with access policies and ensures that all changes can be traced back to the responsible party via an attribute authority. Jia et al. (2021) introduced the K-time Modifiable and Epoch-Based Redactable Blockchain (KERB), which uses a monetary penalty system to control rewriting privileges and penalize malicious behaviour.

Xu et al. (2021) developed a system with a supervisory authority that monitors processes and allows users the ability to propose changes and manage their data. This approach introduces a new variant called Stateful Chameleon Hash with Revocable Subkey, which builds on Policy-Based Chameleon Hash and incorporates a practical Revocable Attribute-Based Encryption (RABE) scheme to achieve a revocable policy-based Chameleon hash.

Panwar, Vishwanathan and Misra (2021) presented a framework for revocable and traceable blockchain rewriting, based on a revocable chameleon hash with an ephemeral trapdoor scheme and revocable CP-ABE. Liu et al. (2021) was the first to discuss replacing traditional hash functions with distributed chameleon hash and integrating distributed key management with a trusted execution environment. Matzutt et al. (2022) introduced parallel jury committees to perform modifications and distribute trapdoor keys among committee members.

Chameleon hash to enable mutability in blockchains is an active area of research for which numerous solutions are proposed. The common goal of these solutions is to calculate collisions in such a way that the changed block or transaction hash or signature of the transaction remains unchanged.

### 5.1.2. Consensus

Consensus-based approaches rely on the collective agreement of the network majority to implement changes to the state of the blockchain. Redaction is achieved through the voting of honest participants. If the majority agrees on the redaction changes, the redaction state is finalized.

*Hard Fork:* A hard fork occurs when a blockchain splits into separate chains, creating a new version that is incompatible with the older version. This split prevents nodes on one fork from interacting with those on the other. The longer chain is usually considered the authentic chain. An early attempt to modify on-chain changes made through a hard fork was the separation of Ethereum and Ethereum Classic following the DAO incident (Falkon, 2018). Other examples include Bitcoin Gold and Bitcoin Cash, which were created through a hard fork from Bitcoin (Webb, 2018). However, hard forks are often viewed as inefficient redaction mechanisms. They do not delete transactions, but rather create two separate versions of the chain, with the older chain's state still accessible to their nodes. Additionally, hard forks require significant resources to execute.

*5.1.2.1. Consensus Voting*

Deuber, Magri and Thyagarajan (2019) proposed a solution based on consensus voting for public and permissionless blockchains. CV Chain extends block header structures with additional fields to record the old state before redaction and creates a connection between the redacted state and the new using two hashes - one for the old state and one for the new state. In this system, miners verify modified blocks and vote on valid revisions within a certain voting period. Revisions will be approved if they receive enough votes. Honest nodes then update their personal copies with the newly applied modifications. Despite its advantages, this solution still faces challenges in terms of scalability and efficiency due to the time required to approve valid redactions.

Marsalek and Zefferer (2019) introduced the concept of a correctable blockchain, which consists of two chains: the standard chain that stores original data and the correction chain that stores correction data. This system uses a consensus voting mechanism to make decisions about corrections in a decentralized manner.

Thyagarajan et al. (2021) proposed a publicly verifiable repair layer that can repair or modify blocks without creating a new chain. This repair layer can be added as an additional layer to any blockchain. The process is initiated by a repair message and uses a voting mechanism. Once the repair request receives enough votes, the repair will be approved.

Consensus-based solutions do not rely on central authority or trust assumptions and are therefore suitable for permissionless networks. However, delays in reaching consensus can cause problems, especially when there are multiple requests for modifications at the same time.

### 5.1.3. Meta Transaction

Meta transactions provide an alternative to heavy cryptographic methods such as chameleon hash and utilize special transactions called meta transactions. Puddu et al. (2017) proposed a solution called μchain, where all modifications on the blockchain are controlled by fiat, enforced by consensus, and are verifiable like normal transactions. Meta transactions introduce a set of possible transaction versions, an active transaction, and a mutability policy. Mutations are governed by policies that define access control measures and the time window during which a transaction remains modifiable. Transactions are grouped into sets consisting of an active version and several inactive alternatives. This solution encrypts all historical versions of transactions and provides the decryption key only for the active record.

### 5.1.4. Pruning

Pruning is the process of removing unnecessary data blocks from the blockchain, specifically targeting older blocks or transactions that are no longer needed. While the original purpose of pruning was to reduce blockchain size and improve efficiency, it also offers benefits for privacy and regulatory compliance.

Matzutt et al. (2020) introduced a solution called CoinPrune for pruning the Bitcoin blockchain. CoinPrune creates snapshots of the UTXO set every 10,000 blocks, containing headers and serialized UTXO sets of the pruned blocks. Miners publicly announce these snapshots, and other miners independently verify their correctness. New nodes use these snapshots for initial synchronization, significantly reducing the amount of data they need to download and process.

Pruning allows for the removal of older blocks under the assumption that they are independent of verifying future transactions. However, this method may come with certain security trade-offs.

### 5.1.5. Local Erasure

In the functionality-preserving local erasure (FPLE) approach suggested by Florian et al. (2019), node operators mark specific parts of transactions for erasure and store them in an erasure database. Erasure is performed locally on individual nodes by adjusting the ScriptPubKey field of transaction outputs. Erased data is replaced with substitute values within the transaction. Redacted versions of transactions (T') are stored in the erasure database. If data has been erased, the stored redacted transaction (T') is used for subsequent operations to verify integrity. However, the approach of local erasure presents a challenge: data may still be stored in some other nodes.

### 5.1.6. Polynomial Functions

A method that uses polynomial functions to organize data segments within each block was proposed by Cheng et al. (2019). This approach employs Lagrange interpolation methods to structure data and gives users varying levels of control over changes, such as adjusting the levels of control over modifications, adjusting difficulty and setting coordination conditions.

Blocks contain headers with the order of the block, information about the finite field of the blockchain, the difficulty of modification and coordinates. The body contains records, each containing data, additional information, padding, and a coordinate. Each block is linked to its predecessor by recording a polynomial function in its header. This function, derived from the coordinates in the block's label fields using Lagrange interpolation, ensures that each block has a unique polynomial representation.

To integrate n data segments into the chain, the process involves identifying the polynomial function of the previous block and including it in the new block's header. A prime number is selected, and modification difficulty is set, both stored in the new block's header. The padding field is left empty. Coordinates for the new block are calculated, and for each data segment, a new record with essential information is created. The new block is then broadcast to complete the process.

Updating a block replaces old data with new data and updates the associated information. Random characters are added to the fill field and a new position is calculated based on a specific formula. This is repeated until the new position meets the difficulty criteria. The new position is recorded and a new record with updated data, information and position is created. The old record is replaced with the new one in the block and the updated block is broadcast.

To delete a block, the order of the blocks is adjusted by changing the order field of the following block, while ensuring that the polynomial function remains unchanged. The modified block is then sent to replace the original one. This method also allows inserting a new block between existing blocks and adjusting the order accordingly to maintain the sequence.

### 5.1.7. RSA

The RSA-based approach for modifying blockchain blocks, proposed by Grigoriev and Shpilrain (2020), It includes both public and private information: a public hash function generates unique codes for each block, while private prime numbers are used to create a large number required for encryption. Each block is divided into three parts: a beginning (prefix), a middle (content) and an end (suffix). The prefix and content are hashed together using the public hash function and a mathematical operation involving the suffix generates the prefix of the next block. A central authority with access to the private key can modify the contents of a block while maintaining the integrity of the chain by adjusting the suffix accordingly.

Although not originally designed for blockchains, Sanitizable signatures and redactable signatures offer methods for selectively modifying signed messages. According to a review by Bilzhause, Pöhls and Samelin (2017), sanitizable signature techniques enable the removal or masking of specific parts of a signed message. In these solutions, particular sections of the message are marked for redaction and replaced with special symbols or placeholders, preserving the original structure. The verifier can then authenticate the remaining content using the signature and public parameters while confirming that the redacted parts were appropriately removed. Similarly, Zhu et al. (2021) proposed the Identity-Based Redactable Signature Scheme (IDRSS). Using the redaction mechanism provided by IDRSS, specific parts of a signed message can be selectively deleted.

### 5.1.8. Data Block Matrix

A data structure known as the data block matrix facilitates the overwriting of blocks with zeros (Kuhn, 2022). This structure resembles a specialized table capable of storing information, with each block containing a hash value linked to neighbouring blocks. The data block matrix is designed to selectively erase specific parts of the information while preserving the integrity of the remaining data. Each piece of information in the table is associated with a hash value, ensuring data integrity. When a portion of the information is erased, it is replaced with zeros. Subsequently, the hash values of the affected row and column are recalculated to maintain consistency across the matrix. This organized and balanced structure ensures that all elements remain correctly aligned.

# 6. IoT-Specific Implementations and Use Cases

The concept of redactable blockchain holds promise for addressing the challenges of data modification and deletion in IoT-based blockchain systems. Our focus will shift to exploring implementations and use cases of redactable blockchain solutions specifically tailored for IoT environments. We have identified implementations that present a wide range of approaches and techniques. Below we provide a detailed overview of notable implementations and examine their specific characteristics and methods.

## 6.1. Chameleon Hash Based

### 6.1.1. Redactable Consortium Blockchain (RCB) for IIOT

The RCB for IIoT, introduced by Huang et al. (2019), employs chameleon hash functions to enable redaction. It integrates Threshold Chameleon Hash (TCH) and Accountable and Sanitizable Chameleon Signatures (ASCS), enabling changes to file and block level signatures. The system includes four entities: Chain Manager, User Sensors, Authorized Sensors, and Judge Sensors. The chain manager initiates the chain, selects authorized sensors, and monitors disputes. User sensors, which are IIoT devices, use ASCS to sign transactions. Authorized Sensors write and redact the blockchain and update signatures, while Judge Sensors resolve disputes about the authenticity of the signature.

The TCH and ASCS algorithms generate system parameters and keys so that transactions can be signed and verified using ASCS.Sign and ASCS.Verify. The chain manager sets up the system and orchestrates membership, with TCH.Hash replacing the roots of the Merkle tree to allow redaction without changing block hashes. Block redaction requires consensus between authorized sensors, supported by smart contracts.

For redaction, authorized sensors use TCH.Forge to collectively generate new randomness for chameleon hashes, enabling blockchain data to be updated. ASCS.Sanitize handles signature sanitization and supports both coarse-grain and fine-grain redaction.

The evaluation results show efficient hash computation and granular data control with TCH and ASCS, although the increased computational complexity of ASCS.SS may affect scalability. Experiments demonstrate redaction efficiency, particularly at small scales or at the file level, with a trade-off between security and efficiency.

### 6.1.2. self-redactable blockchain (SRB)

By Huang et al. (2020), a self-redactable blockchain (SRB) based on a revocable chameleon hash (RCH) was proposed. The framework includes users, miners, and the SRB and allows users to independently perform redactions using ephemeral trapdoors that periodically expire to prevent misuse.

The SRB setup includes the generation of system parameters, keys and ephemeral trapdoors. Keys are randomly generated, and hash computation involves computing chameleon hash and randomness based on system parameters, identity, time, hash key and message. Transactions are authenticated by signing the message, including the transaction ID, last signature and the next owner's public key.

The solution also introduces an RCH-derived revocable chameleon signature (RCS), which enables public redaction of signatures within a specified time period. Both private and public redactions can be performed, with public redaction being available to anyone. Redaction of a transaction involves one private redaction and multiple public redactions.

The SRB performance evaluation shows efficient hashing, verification, and trapdoor generation operations, with features such as trapdoor expiration enhancing security. However, the forging algorithm has lower efficiency, which may affect the overall performance. Computational costs and energy consumption increase with the number of transactions and blocks, leading to scalability challenges. Energy consumption for RCH hashing is higher than SHA-256, indicating inefficiencies.

### 6.1.3. Scalable and redactable blockchain

Huang et al. (2021) introduced a Scalable and Redactable Blockchain (SRB) that leverages two cryptographic protocols: Time Updatable Chameleon Hash (TUCH) and Linkable-and-Redactable Ring Signature (LRRS). TUCH enables spontaneous ring formation and periodic chameleon randomness, while LRRS enables anonymous signing and redaction of signatures. These protocols form the basis of the SRB.

The SRB system includes miners (trapdoor holders) and users. Miners pack signed transactions into blocks using TUCH for hashing, while users sign and publish transactions using LRRS for anonymity and redactability. TUCH handles system setup, key generation, message hashing, verification, collision forging, and hash updates. LRRS manages system setup, key generation, message signing, signature verification, redaction, updating, linking for double-spending detection, and transaction denial.

TUCH and LRRS replace SHA-256 and ECDSA with chameleon hash and malleable signature schemes, TUCH and LRRS serve as the system's hash function and signing scheme. Miners control block redaction through trapdoor keys (tk), with the hash key (hk) recorded on the block. Redaction uses TUCH and LRRS to redact the hash and signature and maintain chain consistency through periodic updates.

The performance evaluation shows that signing is the most time-consuming operation while updating takes the least. TUCH is efficient at hashing, verifying, forging and updating. However, LRRS complexity is affected by the size of the signing ring, which affects signature processing.

### 6.1.4. Industrial Data Management with Redactable Blockchain

Zhang et al. (2021) proposed a trustworthy industrial data management scheme using a redactable blockchain and introduced a double blockchain architecture to separately manage trapdoor transactions and reduce system load. The system includes the system manager, the Key Generation Center (KGC), trapdoor holders and an executor. The system manager initiates the blockchain setup and parameter generation. The KGC generates trapdoors and makes fragments available to trapdoor holders participating in both redactable and supervision blockchains. The executor reconstructs the blockchain after complete trapdoor recovery.

The system consists of several algorithms for setup, trapdoor generation, public key generation, chameleon hash calculation, trapdoor restoration, and data forging. The lifecycle includes an initial setup phase followed by modification cycles where trapdoor generation, key generation, hash calculation, and data forging occur.

Operationally, the off-chain setup phase involves setting system parameters and deploying blockchains, with the KGC distributing trapdoor fragments. In the on-chain setup phase, Trapdoor holders generate the public key for the redactable blockchain. During the data management phase, industry data is stored using the Chameleon hash function, with miners selecting and nodes verifying transactions before storing them in blocks.

Integrity checks are carried out for data processing. When false data is detected, trapdoor holders restore the trapdoor using specific algorithms, modify the data, and broadcast the changes on the redactable blockchain.

The scheme can detect and mitigate malicious behaviours. Performance evaluations indicate reasonable overhead for off-chain operations and effective fault-tolerant trapdoor recovery, assuming there are pre-selected, authorized trapdoor holders.

### 6.1.5. Re-Chain

Re-Chain (J. Zhang et al., 2020) is a rewritable blockchain specifically designed for fixed storage, particularly for edge computing environments. The system is based on a consensus-based mechanism and uses TTCH to effectively regulate rewrite operations. The main entities involved in this system are the cloud, edge nodes, key authority and end devices. The cloud provides storage and computing resources, while edge nodes are responsible for recording transactions with their fixed storage and computing power. The key authority handles key management and initializes the blockchain, and end devices that have limited storage and computing capacity collect data.

The rewriting process in Re-Chain is controlled by a consortium of edge nodes, ensuring validation in a similar way to ordinary transactions. TTCH is based on the public coin Chameleon Hash by Khalili, Dakhilalian and Susilo (2020) and includes setting up bilinear group parameters and generating keys. The construction of TTCH includes the generation of keys as well as the computing hash, verification and signing processes.

An edge node initiates the process by proposing a block containing transactions. This node verifies the block ID, rewrites old transactions if necessary, and digitally signs the block before broadcasting the proposal. Other edge nodes then validate the proposal's signature and ID. once validated, these nodes sign the Merkle root's child hashes and issue credentials to the proposal node. The proposal node collects these credentials, computes collisions to ensure integrity, calculates a new randomness to enable verification, and broadcasts it to the network. Finally, validators receive the new randomness, validate the Merkle root hash, and rewrite the original block with the new validated one, thereby achieving consensus.

The proposed consensus mechanism in Re-Chain ensures secure and controllable rewriting of blockchain data. Experimental validation shows that the system performs acceptable at medium scales, although performance at larger scales remains uncertain based on the experiments conducted.

### 6.1.6. Secure Federated Learning in Industrial Internet of Things

Wei et al. (2022) proposed a novel chameleon hash scheme with a changeable trapdoor (CHCT) to secure federated learning in Industrial IoT environments by storing and sharing aggregated models on the blockchain. The approach includes limiting trapdoor use and allowing flexible expiration periods chosen by data owners.

In the processing of Federated Learning (FL) with Redactable Blockchain, agents receive training models, encrypt them, and store them in the cloud. Agents broadcast transactions and store access links in the blockchain. If a privacy risk is detected, data is processed in the cloud to create a new access link, and the agent calculates a valid chameleon hash collision to modify the old access link in the blockchain.

The system includes three main entities: data owners, data modifiers and supervisors. Data owners calculate the initial chameleon hash tuple and send the trapdoor to the data modifier. Data modifiers, with the permission of the data owners, modify the data and calculate a chameleon hash collision with a supervisor-approved trapdoor update. Supervisors monitor the process, record public information, verify the Chameleon hash tuple, and ensure the identity of the data owner in case a revocation of a prior authorization is required.

The operation of CHCT involves multiple steps. Trapdoor Key Management facilitates secure sharing of trapdoor keys between entities. Trapdoor Update Determination sets the trapdoor update parameter based on collected points. Collision Detection identifies collisions for chameleon hash tuples to facilitate trapdoor updates, while Collision Verification verifies these collisions and updates related parameters. Trapdoor Abolishment securely abolishes old trapdoors and generates new ones. Commitment Opening verifies entity identity and updates chameleon hash tuples. Updating the trapdoor upon detecting a collision enhances collision resistance and resilience against collision attacks.

Duan et al. (2023) proposes the Policy-Based Chameleon Hash with Black-Box Traceability (PCHT), a scheme intended to improve redactability and traceability in blockchain systems. PCHT uses attribute-based encryption and collision-resistant chameleon hash to identify unauthorized changes involving entities such as the IoT data owner, authority, and modifier. Although this scheme is primarily a new chameleon hash proposal, it provides a foundation for secure and traceable data management in IoT environments and supports blockchain-based transactions and modifications with robust tracing capabilities.

### 6.1.7. Policy-hidden Fine-grained Redactable Blockchain (PFRB)

Guo et al. (2023) authors present a policy-hidden fine-grained redactable blockchain (PFRB) that is intended to improve security and data protection in IoT systems. PFRB leverages chameleon hash functions and secret sharing based on Newton's interpolation formulas to enable fine-grained redactions controlled by hidden policies.

The system architecture includes several key entities: authorities, transaction owners, transaction modifiers and blockchain participants. These entities work together within the system framework, with authorities initiating system processes and transaction owners adding data to the blockchain, while transaction modifiers are tasked with modifying existing transactions based on predefined policies.

The PFRB framework builds on an existing scheme (RBDS22) (Ma et al., 2022) and includes improvements to policy handling and redaction control. The main technical components include Setup Algorithm, RKGen, ModSetup, AuthSetup, ModKeyGen, Hash Function, Verify and Adapt. These algorithms collectively facilitate system initialization, key generation, policy enforcement, and transaction verification.

The performance evaluation highlights several strengths of the proposed solution. Thanks to efficient linear secret sharing matrices, it has superior key generation efficiency and takes significantly less time compared to traditional methods. Hashing efficiency is also improved, showing linear growth instead of the exponential increase seen with traditional methods due to the use of polynomial functions. Verification time remains comparable to traditional methods, with response time increasing linearly. However, both the proposed solution and traditional approaches experience an exponential increase in adaptation time as the number of policies increases.

### 6.1.8. Auditable Redactable Blockchain and ACHR

Shao et al. (2023) propose an IoT-focused system using the Auditable Chameleon Hash with Revocability (ACHR) scheme. The solution enables IoT user devices to independently manage their on-chain transactions while providing "verify-then-modify" auditability and mandatory revocability.

The key entities in this system include user devices, auditors, and the ledger. The ACHR scheme consists of algorithms for setup, key generation, hashing, commitment, proof, ciphertext issuance, hash and pre-signature adaptation, hash verification, auditing trapdoor extraction, and user trapdoor revocation.

ACHR uses two Chameleon Hash functions for user and auditor key generation, with a two-layer CH allowing external trapdoor holders to revoke rewriting without altering the hash value. Privacy is maintained through zero-knowledge proofs and encrypted presignatures. Modified transactions include auditing proofs for verification and secret extraction by auditors to ensure accountability.

The evaluation reveals challenges such as the time and memory overhead of extract, adapt, and revoke operations, as well as potential impact on throughput with larger input message sizes. Key length dependency also impacts performance.

### 6.1.9. Redactable Blockchain-Based Data Management Scheme for Agricultural Product Traceability

Yang et al. (2024) proposes a hybrid on-chain and off-chain model to support threshold editing and mitigate single-point-of-failure risks. The system includes three main entities: System Administrator (SA), Authorized Agricultural Entities (AAEs), and the Cloud Server (CS). The SA is responsible for setting up the system, managing users, and configuring the blockchain but does not participate in block creation or modifications. AAEs are agricultural enterprises and regulatory agencies that participate in traceability and blockchain editing. The CS stores agricultural product traceability data (APTD) and helps distribute the storage load.

The solution uses an Attribute-Based Proxy Re-Encryption (ABPRE) algorithm to manage data access rights. The process begins with the SA creating system parameters and setting up the blockchain. AAEs register to receive encryption and editing keys. Farmers and IoT devices encrypt APTD with AES, while access control uses attribute-based methods. Digital digests are uploaded to cloud storage and blocks are validated and confirmed by network entities.

Scenarios during the data editing phase include correcting data errors and changing access permissions using proxy re-encryption. The data provider initiates these requests. During editing, AAEs check validity, compute collisions for integrity, and modify blocks as necessary. An accountability mechanism ensures security by detecting and managing malicious actions during editing.

The query process involves searching the blockchain, retrieving ciphertext, and decrypting data. The scheme demonstrates efficient data uploading and editing processes, outperforming alternative schemes like ASCS and PCHBA with 42% faster block generation and 29.3% faster block editing with 125 nodes. The accountability mechanism is effective even with malicious entities, and adjusting the DTCH threshold can optimize speed.

## 6.2. Pruning Based
### 6.2.1. MOF-BC (Memory Optimized and Flexible BlockChain)

Dorri, Kanhere and Jurdak (2018) proposed MOF-BC, which introduces memory optimization strategies such as User-Initiated Memory Optimization (UIMO), Service Provider-Initiated Memory Optimization (SIMO), and Network-Initiated Memory Optimization (NIMO). These strategies allow for flexible transaction storage and block hashes are calculated based on transaction hashes rather than their contents.

MOF-BC uses multiple agents to manage different aspects of the blockchain:

- Summary Manager Agent (SMA): Consolidates multiple transactions into one.
- Reward Manager Agent (RMA): Calculates rewards for memory optimization.

- Storage Manager Agent (StMA): Collects storage fees and distributes them to miners.
- Patrol Agent (PA): Monitors miners' storage resources and verifies claims.
- Blackboard Manager Agent (BMA): Manages essential information.
- Service Agent (SerA): Processes transaction removals and updates the blockchain.
- Search Agent (SA): Identifies specific transactions for processing.

MOF-BC includes a storage fee in transaction fees to incentivize nodes to store blockchain data. UIMO uses a Generator Verifier (GV) to create transactions, eliminating the need for multiple key management. Users can remove stored transactions by proving ownership through hashes used to generate the GV. The PA verifies removal claims, and miners process removals in batches during a periodic Cleaning Period (CP), rewarding users for memory optimization.

MOF-BC allows for transaction summarization and aging to compress stored data. Summarization consolidates multiple transactions, and aging replaces original transactions with compressed versions. The BMA ensures the accessibility and integrity of redirected transactions.

SIMO enables Service Providers (SPs) to initiate memory optimization without user intervention and use GV-PK+ and GVS to manage requests. NIMO shifts memory optimization to the network, reducing user and SP involvement. Transactions are categorized as do not store, temporary, permanent, or summarizable, with the SMA managing summarization.

When transactions are removed, their content is deleted, but IDs are retained. For ledger removal, only genesis transaction identifiers are needed. The CP determines batch removal timing, managed by the SerA. Effective memory optimization significantly reduces the blockchain's memory footprint, with larger CP values improving throughput and scalability, though they may increase memory waste and costs for miners.

### 6.2.2. LiTiChain

Pyoung and Baek (2020) introduced LiTiChain, a pruning-based blockchain solution that incorporates blocks with finite lifetimes, enabling the safe removal of outdated transactions and blocks. This approach improves scalability and serves as a deletion solution in blockchain systems.

LiTiChain's design involves two primary graphs: one for the expiry order of lifetimes (Endtime Ordering Graph, EOG) and another for the block creation order (Arrival Ordering Graph, AOG). The IoT system include IoT devices and edge servers communicating via a peer-to-peer network. Edge servers handle the processing, storage, and control of IoT data transactions using permissioned blockchain technology.

Block lifetime in LiTiChain is defined from block creation to the latest transaction endtime. Blocks with expired transactions are eligible for deletion. The EOG ensures chain connectivity by creating a directed edge from each new block to an existing block based on the endtime attribute, forming a tree structure with the genesis block as the root node.

The AOG extends the EOG by linearly linking blocks based on their creation order, providing a conventional blockchain structure. Each block header contains ParentBlockHash (EOG) and PreviousBlockHash (AOG). Blocks must persist until dependent blocks are validated, incurring retention costs (RC). To manage RC, edges from the AOG are constrained using the K parameter.

Transactions are categorized by lifetime properties:

- Fixed: Known lifetime at creation.
- Permanent: Infinite lifetime.
- Indefinite: Lifetime not decided at creation, can expire anytime.

Deletion of indefinite transactions triggers block renewal, creating new blocks marked for deletion after a period defined by parameter D. Bogus or Mock blocks are periodically generated to maintain blockchain height, with renewal or deletion based on average maximum heights.

### 6.3. Other
### 6.3.1. Privacy-Preserving Redactable Blockchain

Ren, Cai, and Hu (2021) have introduced a transaction-level blockchain redaction method that ensures user privacy using threshold ring signatures and symmetric encryption. The solution includes four key algorithms:

- validateReq: Validates redaction requests.
- validateCandTx: Validates candidate transactions for redaction.
- validateChain: Ensures chain integrity.
- validateBlock: Ensures individual block validity.

Regular transactions contain encrypted data and ring signatures. Users or miners can propose redaction requests for specific transactions, involving information disclosure and new data generation. Miners validate these requests and participate in redaction by voting and generating threshold ring signatures. Any network participant can verify proposed redactions.

The system efficiently redacts transactions, with an overhead ratio of 76% to 84%, demonstrating higher efficiency compared to other deletable chain. While it supports user validation of operations, generating threshold ring signatures and encrypting data may require more computational resources.

## 6.3.2. Redactable Blockchain-Assisted Secure Data Aggregation Scheme for Fog-Enabled IOFT

Mishra et al. (2023) presents a three-tier architecture designed to enable efficient and secure two-tier data aggregation in fog-enabled Internet of Farming Things (IoFT) environments. This architecture integrates a redactable blockchain framework into the fog layer.

At its core, the proposed architecture consists of three layers: the IoFT layer, the fog layer and the cloud server. The IoFT layer includes IoT devices deployed in agricultural fields to monitor crop status and transmit data to the Access Control Center (ACC) via fog servers. The Fog Layer acts as an intermediary, facilitating local data aggregation while leveraging cryptographic techniques such as the Paillier cryptosystem and the certificateless signature scheme for increased security. Additionally, the Fog Layer uses a committee-endorsing mechanism to effectively manage blockchain operations. Meanwhile, the cloud server hosts the ACC for real-time data analysis and the Trusted Key Authority (TKA), which is responsible for system initialization.

As data flows in from IoFT devices, the fog layer handles the critical tasks of data validation and aggregation. Fog servers authenticate incoming data to ensure its integrity before calculating tag values and sending the aggregated data to specific nodes for further processing. A randomly selected leader node then generates a block containing the aggregated data and the associated validation values. After successful verification, this block is integrated into the blockchain and the ledger is updated.

Fog servers initiate block modifications by selecting the relevant block within the blockchain and replacing outdated data with updated information. The modified block is then broadcasted for validation, with stringent criteria ensuring compliance with expiration times and redaction policies. Through a proof-of-work (PoW) process, other nodes validate the modified block, updating the blockchain if consensus is achieved.

The paper also highlights the need for a smart consensus algorithm, although it does not propose a specific solution. Additionally, it mentions the importance of incentivizing all involved entities but does not delve into designing an appropriate incentive mechanism.

| Author and Year | Blockchain Scheme Identification | Redaction Schema and Methods | Chain Growth and Block Consistency | Stored On-Chain Data | Off-Chain Stored Data | Setting | Blockchain Type | Redaction Type | Modifier | Centralized Modifier | Pre-Selected Modifier | Modification | Deletion | Accountability | Encryption |
|---|---|---|---|---|---|---|---|---|---|---|---|---|---|---|---|
| Dorri, Kanhere and Jurdak (2018) | Memory Optimized and Flexible Blockchain (MOF-BC) | UIMO, SIMO, NIMO | ✓ | IoT Device Generated Data | - | IOT | Consortium | transaction level | - | ✗ | - | ✗ | ✓ | - | ✓ |
| Huang et al. (2019) | Redactable Consortium Blockchain (RCB) | TCH and ASCS | ✓ | IoT Device Generated Data | - | Industrial Internet of Things | Consortium | transaction level and block level redactions | Authorized sensors | ✗ | ✓ | ✓ | ✓ | ✓ | - |
| Pyoung and Baek (2020) | LiTiChain | Blocks with Finite Lifetime | ✓ | IoT Device Generated Data | - | Edge-based IoT | Consortium | block level | - | ✗ | - | ✗ | ✓ | - | - |
| Huang et al. (2020) | Self-Redactable Blockchain (SRB) | Revocable Chameleon Hash (RCH) | ✓ | Not Explicitly Mentioned | - | IOT | - | block level | Authorized User | ✗ | ✓ | ✓ | ✓ | ✓ | ✓ |
| Zhang et al. (2020) | Re-Chain | TTCH | ✓ | IoT Device Generated Data | Backup of Overwritten Data | Edge-based IoT | Consortium | block level | Edge Nodes | ✗ | - | ✓ | ✓ | ✓ | - |
| Huang et al. (2021) | Scalable and Redactable Blockchain | TUCH and LRRS | ✓ | IoT Device Generated Data | - | IOT | Permissionless | block level | Miners | ✗ | - | ✓ | ✓ | ✓ | ✓ |
| Zhang et al. (2021) | Redactable Blockchain and Supervision Blockchain | CH | ✓ | Redactable Chain: IoT Device Generated Data. Supervision Chain: Verification Data and Trapdoor Fragments | - | Industrial Internet of Things | Consortium | block level | Trapdoor Holders | ✗ | ✓ | ✓ | ✓ | ✓ | - |
| Ren, Cai and Hu (2021) | Privacy-Preserving Redactable Blockchain | Threshold Ring Signature, Symmetric Encryption | ✓ | Not Explicitly Mentioned | - | IOT | Consortium | transaction level | Miners | ✗ | - | ✓ | ✓ | ✓ | ✓ |
| Wei et al. (2022) | Redactable Medical Blockchain (RMB) | Chameleon Hash with Changeable Trapdoor (CHCT) | ✓ | Hash to the Model, Metadata | Actual Federated Learning Model | Industrial Internet of Things | Consortium | transaction level | Data Modifier with Owner's Permission. | ✗ | - | ✓ | ✓ | ✓ | ✓ |
| Duan et al. (2023) | - | ABET and CHET | ✓ | IoT Device Generated Data | - | IOT | consortium | transaction level | Transaction Modifiers | ✗ | ✓ | ✓ | ✓ | ✓ | ✓ |
| Guo et al. (2023) | Policy-Hidden Fine-Grained Redactable Blockchain (PFRB) | CH and MA-ABE and Secret Sharing | ✓ | Policy Contents and Transactions | - | IOT | Permissionless | transaction level | Transaction Modifier | ✗ | - | ✓ | ✓ | ✓ | ✓ |
| Mishra et al. (2023) | - | Consensus | ✓ | Encrypted IoT Device Collected Data | Aggregated Data from Fog Servers | Internet of Farming Things | Consortium | block level | Leader Fog Node | ✗ | ✓ | ✓ | ✓ | ✓ | ✓ |
| Shao et al. (2023) | Auditable Redactable Blockchain, ACHR | ACHR | ✓ | IoT Device Generated Data | - | IOT | Consortium | transaction level | User Devices | ✗ | - | ✓ | ✓ | ✓ | ✓ |
| Yang et al. (2024) | - | DTCH | ✓ | Data Summaries and Access Control, Hashes of Data | Actual Data | Agricultural Product Traceability | Consortium | block level | Authorized Entities | ✗ | - | ✓ | ✓ | ✓ | ✓ |

Figure 1: Comparison of Redactable Blockchain Implementation for IoT Applications. Detailed data is available in the high-resolution version on Zenodo at: https://doi.org/10.5281/zenodo.11847110

# 7. Current Challenges and Future Research Directions

Research on redactable blockchain design is still in its early stages and is expected to attract more attention in the near future. The concept of redactable blockchain is promising but remains largely theoretical, with no prominent real-life implementations yet. This section describes the current challenges faced by redactable blockchain solutions and suggests future research directions.

Redactable blockchain solutions are still new and will continue to be developed. Most of these solutions leverage permissioned and consortium blockchain networks. However, implementing redactable solutions in permissionless blockchains remains a major challenge due to their open, large, and unrestricted nature. Redaction itself poses problems as it violates the blockchain's fundamental immutability property, raising concerns about the integrity and trustworthiness of the system.

The complexity of implementing some ideas may result in changed transactions not being verified correctly or breaking consistency. Consensus delays continue to be a problem and can lead to weaknesses in consensus-based redaction processes. Many solutions present scalability and efficiency problems. While pruning methods can improve scalability, they can also lead to more vulnerability to attacks and problems with proof of validity due to the consistent removal of large numbers of blocks.

Centralized control is another concern, as some solutions may suffer from this due to the presence of pre-selected or authorized modifiers. This approach can enhance security but reduces flexibility and decentralization, which are core attributes of blockchain technology. Moreover, some solutions may avoid accountability by preserving anonymity and hiding identities, which can be misused by authorities with modification powers. Measures are being implemented to prevent false data modifications, but the risk of misuse remains.

Computational requirements present additional challenges. Methods such as proof of work (PoW) may not be suitable for IoT devices with limited resources. Chameleon hash methods, while efficiently used in permissioned and consortium blockchains, they encounter issues related to editing privileges and key exposure. These solutions are less effective in permissionless chains due to key management and sharing issues. Although almost all chameleon hash-based solutions have collision resistance properties and maintain block consistency and chain growth even after redaction, they may not be suitable for IoT devices with limited computing power.

Redactable blockchains suffer from similar challenges as traditional blockchains, where performance and scalability might not match the requirements of IoT systems. These systems can be slower and costlier to adapt for IoT, requiring traditional blockchain strategies to address the scalability and performance gap. In scenarios where the ledger grows with more IoT data, causing slower operations and increased storage requirements, pruning solutions can be beneficial by removing older blocks and keeping the chain at a healthy size. Centralized redaction can improve performance but may lead to a loss of decentralization. Redactable blockchains for IoT might perform well with medium-sized networks but may not be suitable for larger networks.

As quantum computers mature, some mathematical problems will become solvable and pose a threat to current cryptographic techniques. Recent advances have been made in quantum-resistant chameleon hash methods to address these challenges (Wang et al., 2024; Wu, Ke and Du, 2021; Thanalakshmi et al., 2021), and further research is expected in this area.

Overall, although redactable blockchains offer promising solutions for various applications, there are significant challenges that need to be overcome. Future research should focus on developing more robust and efficient redactable blockchain designs. Specific areas of interest include scalability, security improvement, efficiency and resource optimization for IoT scenarios. Decentralization and exploration of permissionless networks are also critical areas for future work. In IoT, redactable blockchains can be used for data removal to meet privacy requirements, optimize storage usage in edge devices, update access controls, and manage identities. Research could aim to make these solutions practical for real-world IoT applications.

## 8. Conclusion

In conclusion, integrating blockchain technology into the Internet of Things (IoT) offers promising solutions to improve data security and integrity. However, the immutability of blockchain presents challenges in complying with data protection laws that require data redaction. Although redactable blockchain solutions show potential, they are still in their early stages and have various limitations. Future research should focus on improving scalability, security, and real-world implementation to fully harness the benefits of redactable blockchains in IoT applications. Despite these challenges, progress in redactable blockchains is promising and further research and development is expected to advance their development in the future.

## 9. Acknowledgments

I acknowledge the use of OpenAI's GPT-3.5 (Open AI, https://chat.openai.com) to paraphrase technical content and improve the readability of this manuscript.

I would also like to thank Rohit Raj for his contribution to the section titled "Data Protection Laws and Regulatory Requirements."